\documentclass[british,english]{article}
\usepackage[T1]{fontenc}
\usepackage[latin9]{inputenc}
\usepackage{geometry}
\usepackage{color}
\usepackage{amsmath}
\usepackage{amssymb}
\usepackage[authoryear]{natbib}

\usepackage{expl3,xparse}
\usepackage{subcaption}
\usepackage{graphicx}

\ExplSyntaxOn
\keys_define:nn { gerry/multifig }
 {
  align .tl_set:N        = \l_gerry_multifig_align_tl,
  scale .tl_set:N        = \l_gerry_multifig_scale_tl,
  innerscale .tl_set:N   = \l_gerry_multifig_innerscale_tl,
  caption .tl_set:N      = \l_gerry_multifig_caption_tl,
  shortcaption .tl_set:N = \l_gerry_multifig_shortcaption_tl,
  label .tl_set:N        = \l_gerry_multifig_label_tl,
  align .initial:n       = c,
  scale .initial:n       = 1,
  innerscale .initial:n  = 0.95,
 }
\NewDocumentEnvironment{multifig}{O{htp}}
 {
  \begin{figure}[#1]
 }
 {
  \end{figure}
 }

\NewDocumentCommand{\shortfig}{mm}
 {
  \group_begin:
  \keys_set:nn { gerry/multifig } { #2 }
  \begin{subfigure}[\l_gerry_multifig_align_tl]{\l_gerry_multifig_scale_tl\textwidth}
  \centering
  \includegraphics[width=\l_gerry_multifig_innerscale_tl\textwidth]{#1}
  \tl_if_empty:NTF \l_gerry_multifig_shortcaption_tl
   {
    \caption{\l_gerry_multifig_caption_tl}
   }
   {
    \caption[\l_gerry_multifig_shortcaption_tl]{\l_gerry_multifig_caption_tl}
   }
  \tl_if_empty:NF \l_gerry_multifig_label_tl { \label{\l_gerry_multifig_label_tl} }
  \end{subfigure}
  \group_end:
 }
\ExplSyntaxOff

\newcommand{\M}[1]{\ensuremath{M=#1}}
\newcommand{\mut}{\ensuremath{\mu_{\theta}}}
\newcommand{\El}[1]{\ensuremath{L=#1}}

\makeatletter

\providecommand{\tabularnewline}{\\}

\makeatother

\usepackage{algpseudocode}
\usepackage{algorithm}

\usepackage{babel}
\begin{document}

\title{Bootstrapped synthetic likelihood}

\author{Richard G. Everitt}
\maketitle
\begin{abstract}
Approximate Bayesian computation (ABC) and synthetic
likelihood (SL) techniques have enabled the use of Bayesian inference
for models that may be simulated, but for which the likelihood cannot be evaluated pointwise at values of an unknown parameter
$\theta$. The main idea in ABC and SL is to, for different values
of $\theta$ (usually chosen using a Monte Carlo algorithm), build
estimates of the likelihood based on simulations from the model conditional
on $\theta$. The quality of these estimates determines the efficiency
of an ABC/SL algorithm. In standard ABC/SL, the only means to improve
an estimated likelihood at $\theta$ is to simulate more times from
the model conditional on $\theta$, which is infeasible in cases where
the simulator is computationally expensive. In this paper we describe
how to use bootstrapping as a means for improving SL
estimates whilst using fewer simulations from the model, and also
investigate its use in ABC. Further, we investigate the use of the
bag of little bootstraps as a means for applying this approach to
large datasets, yielding Monte Carlo algorithms that accurately
approximate posterior distributions whilst only simulating subsamples
of the full data. Examples of the approach applied to i.i.d., temporal
and spatial data are given.

\end{abstract}

\section{Introduction}\label{sec:Introduction}

This paper is concerned with performing Bayesian inference for parameter
$\theta$ conditional on data $y$ (consisting of $N$ data points)
using the prior $p\left(\theta\right)$ and likelihood $L_{\theta}\left(y\right)$,
in situations where the likelihood cannot be evaluated pointwise at
$\theta$ but where it is possible to simulate from $L_{\theta}$
for each $\theta$. Such a choice of $L_{\theta}$ is sometimes referred
to as an ``implicit'' model; exact Bayesian inference is rarely
possible in this setting. The most common approach to inference in
this setting is approximate Bayesian computation (ABC): a technique
for approximate Bayesian inference originally introduced in the population
genetics literature \citep{Pritchard1999,Beaumont2002}, but which
is now used for a wide range of applications including ecology \citep{VanderVaart2015},
cosmology \citep{Akeret2015,Jennings2017}, epidemiology \citep{Kypraios2017}
and econometrics \citep{Martin2017}. ABC has also been used for inference
for models where that are intractable due to the presence of a partition
function \citep{Grelaud2009,Everitt2012}, such as undirected graphical
models.

The key idea in ABC is to approximate the likelihood at each $\theta$
based on simulations from $L_{\theta}$. Usually a non-parametric
kernel estimator of the likelihood is employed, with a bandwidth parameter
$\epsilon$ used to trade off properties of the estimator: for
$\epsilon=0$ the estimator has no bias, but a very high variance,
with the bias increasing and the variance reducing if a larger $\epsilon$
is chosen. In cases where a sufficient (with respect to $\theta$)
vector of statistics $S_{N}$ is available, ABC uses instead an approximation
to the distribution $f_{\theta}$ of $S_{N}$ given $\theta$ in
order to create an approximation with a lower variance. In practice
it is usually only possible to choose a near-sufficient vector of ``summary''
statistics, introducing a further approximation. An alternative approach,
the focus of this paper, is to use a Gaussian approximation \citep{Wood2010f} $\bar{f}_{\theta}$ on the summary statistic
likelihood, with mean $\mu_{\theta}$ and covariance $\Sigma_{\theta}$. This approach
is known as ``synthetic likelihood'' (SL), with $\bar{f}_{\theta}$
being estimated by
\begin{equation}
\widehat{f}_{\theta}=\mathcal{N}\left(\cdot\mid\widehat{\mu}_{\theta},\widehat{\Sigma}_{\theta}\right),\label{eq:sl_llhd}
\end{equation}
where 
\begin{equation}
\widehat{\mu}_{\theta}=\frac{1}{M}\sum_{m=1}^{M}S_{N}^{(m)},\label{eq:sl_mu}
\end{equation}
\begin{equation}
\hat{\Sigma}_{\theta}=\hat{\mathbb{V}}\left[S_{N}^{(1:M)}\right]:=\frac{ss^{T}}{M-1},\label{eq:sl_sigma}
\end{equation}
with $s=\left(S_{N}^{(1)}-\widehat{\mu}_{M,\theta},...,S_{N}^{(M)}-\widehat{\mu}_{M,\theta}\right)$
and $S_{N}^{(m)}$ is the summary statistic vector found from $x^{(m)}\sim L_{\theta}$
for $1\leq m\leq M$ for some $M$ (thus $\hat{\mathbb{V}}$ denotes
taking the sample variance). The approximate likelihood $\widehat{f}_{\theta}$ is evaluated at the statistic vector $S(y)$ of the observed data $y$.  Clearly this approach only provides a
good approximation when the true summary statistic likelihood is approximately
Gaussian, but in practice this occurs in a wide range of applications.
\foreignlanguage{british}{\citet{Wood2010f} applies }this method
in a setting where the summary statistics are regression coefficients
(whose distribution is approximately Gaussian), and recommends transforming
$S_{N}$ for cases where the Gaussian assumption is not appropriate
the original parameterisation.

The SL approximation, using the estimate $\widehat{f}_{\theta}$ may
be used within an MCMC \foreignlanguage{british}{\citep{Wood2010f},
importance sampling or sequential Monte Carlo (SMC) \citep{Everitt2017}
(or Bayesian optimisation \citep{Gutmann2016}) algorithm for exploring
the parameter space.} Monte Carlo methods where the likelihood is
estimated rather than known exactly have been much studied in recent
years. If $\widehat{f}_{\theta}$ were an unbiased estimate of $\bar{f}_{\theta}$, SL would result in
an instance of a pseudo-marginal method \citep{Andrieu2009}, of which
ABC is also a special case. For a pseudo-marginal MCMC algorithm,
the target distribution of $\theta$ is precisely the same as it is
when the exact likelihood is used no matter the variance of $\widehat{f}_{\theta}$,
although a larger variance results in higher variance estimates from
the MCMC output \citep{Andrieu2014a}. However the estimate in equation
\ref{eq:sl_llhd} is biased, thus Monte Carlo methods using
$\widehat{f}_{\theta}$ are a particular case of ``noisy'' Monte
Carlo methods \citep{Alquier2016}, in which the exact target is not
obtained. For such algorithms, under certain conditions, it is possible
to show that the target of the noisy algorithm (using $\widehat{f}_{\theta}$)
converges to the target of the corresponding exact algorithm (i.e.
the ``ideal'' algorithm that uses $\bar{f}_{\theta}$). In the case
of SL, we have such a result as $M\rightarrow\infty$. Thus in practice,
an increased value for $M$ will result in reduced bias and variance
of estimates from the SL-MCMC algorithm, although \citet{Price2017}
find empirically that $M$ does not usually need to be very large
in order that the bias in SL-MCMC is low. \citet{Price2017} also
introduce an unbiased estimator of $\bar{f}_{\theta}$ yielding a
variant of SL-MCMC that has exactly the correct target no matter the
choice of $M$. However, empirically they find that the results are
not significantly improved over standard SL-MCMC.

\citet{Price2017,Everitt2017} find that SL often outperforms ABC,
and is easier to tune, even in some cases where the distribution of
the summary statistics is clearly not Gaussian. However, SL can be
expensive to implement for simulators that have a high computational
cost, since the simulator needs to be run $M$ times for each $\theta$
that is visited. Even if the bias is often low for relatively small
values of $M$, \citet{Price2017} find empirically that for small
$M$ the efficiency of SL-MCMC is poor since the variance of the likelihood
estimator is prohibitively large. In some situations, it may be possible
to exploit the embarrassingly parallel nature of SL and run the $M$
simulators in parallel. However, this is not always possible, in cases
where we wish to use parallelism to explore multiple $\theta$ points
simultaneously, or when a single run of the simulator itself requires
parallel computing.

In addition to running the simulator $M$ times, SL can be costly
when the size $N$ of the data is large (sometimes known as ``tall
data''). In this paper we introduce an approach to using SL where
only subsets of data of size $n\ll N$ are simulated. Using SL (or
ABC) is appealing for tall data, since prior to running an inference
algorithm the dimensionality of the data is reduced by taking a lower
dimensional summary statistic vector. The method then never uses the
full data; the only point in the algorithm that scales with $N$ is
the simulation from $l_{\theta}$. In this paper we show that in some cases this
requirement can be removed, since it is possible to accurately
approximate the posterior using only simulations of size $n$. One
striking result in the literature on Monte Carlo methods for tall
data \citep{Bardenet2017} is that previous methods that use subsamples
of size $n$ (all outside of the ABC/SL context), need to be run for
$N/n$ times as long in order to give the same accuracy as an algorithm
that runs on the full data (with the exception of \citet{Pollock2016}). We show empirically that our approach does not appear to have this requirement (as seen in section
\ref{subsec:Ising-model}).

In summary, this paper investigates methods that provide likelihood
estimates at (sometimes substantially) lower simulation cost, through
reducing the number of simulations $M$ needed from the likelihood,
and also in some cases the size of each simulation from $N$ to $n$.

\section{Methodology}\label{sec:Methodology}

This paper investigates using the bootstrap \citep{Efron1979} as
a means for approximating $f_{\theta}$ using fewer simulations from
the likelihood. Further, we consider the case where the likelihood
is expensive due to its consisting of a large number $N$ of data
points. In this case we investigate the use of the bag of little bootstraps
(BLB) \citep{Kleiner2014} as a means for approximating $f_{\theta}$
that involves simulating only subsets of the full data. This section
gives an overview of the paper, and describes its relationship to
previous work.

To estimate the synthetic likelihood, we must estimate the functions
$\mu_{\theta}$ and $\Sigma_{\theta}$ of $\theta$. The standard
SL approach is to estimate $\mu_{\theta}$ and $\Sigma_{\theta}$
independently for each $\theta$. \citet{Meeds2014a} present an alternative
in which the variance of estimates is lowered by using a Gaussian
process model of each function. For $\Sigma_{\theta}$, this requires
introducing an approximation by modelling only the diagonal of the
matrix. In this paper we remove this requirement by using bootstrap
estimators of $\Sigma_{\theta}$ which we find empirically to have
a lower variance than the raw estimates in equation \ref{eq:sl_sigma}.
\citet{An2016} uses the Graphical Lasso as an alternative approach
to yield low variance estimates of $\Sigma_{\theta}$.

Section \ref{subsec:Bootstrapped-synthetic-likelihoo} describes how
to use bootstrapping to estimate a synthetic likelihood, outlines
some conditions under which this is possible, and describes how to
implement the approach in a computationally efficient way when the
approximation is used within a Monte Carlo algorithm. The use of the
bootstrap in this context has not been considered previously, although
resampling-based ideas have previously been used in the ABC literature
\citep{Peters2010u,Buzbas2015,Vo2015,Zhu2016a}. The only directly
related work to this paper is that of \citet{Buzbas2015}, in which
a small number of simulations from the likelihood are used to construct
an approximate likelihood. A simulation from this approximate likelihood is given by a weighted resampling
of the existing simulations. This approach may be seen as a relatively
crude non-parametric estimator of the likelihood, which we may expect
to be improved using a more sophisticated estimator, such as Gaussian
processes \citep{Wilkinson2014} (also used outside of the ABC context
in \citet{Drovandi2015a}) or neural density estimators \citep{Papamakarios2016}.
Our approach differs in that we use a (parametric) conditional Gaussian
model of the likelihood, and use bootstrapping to estimate its variance.
The method is applicable in any model for which a bootstrapping procedure
is available. We give suggestions for temporal and spatial models
in section \ref{subsec:Temporal-and-spatial}.

In section \ref{subsec:Synthetic-likelihood-with} we extend the method
to cases where a single simulation requires simulating a large number
of data points. Here the BLB is used, leading to a cost that is independent
of the size of the data with little loss of accuracy. Again we describe
how this method may be used efficiently within a Monte Carlo algorithm.

Low variance estimators of $\mu_{\theta}$ cannot be found using the
bootstrap. Therefore we use a variant on the regression ideas in \citet{Meeds2014a}
to estimate $\mu_{\theta}$. Section \ref{subsec:Regression-for-estimates}
describes the approach in full, in which $\mu_{\theta}$ is estimated
via regression and $\Sigma_{\theta}$ is approximated by bootstrap
estimates.

Empirical results for each approach are given in section \ref{sec:Empirical-results},
where we begin by studying a toy example with independent data in
section \ref{subsec:Toy-example} in order to establish the behaviour
of the methods on an example where the ground truth is known. This
is followed by temporal data (from the Lotka-Volterra model) in section
\ref{subsec:Lotka-Volterra-model}, where the summary statistic vector
is 9-dimensional, where the likelihood is difficult to estimate, and
where the summary statistic distribution and posterior are not close
to Gaussian. Finally we study spatial data (from the Ising model)
in section \ref{subsec:Ising-model}, where we focus on using the
BLB to obtain a good approximation to the true posterior in a tall
data setting. Sections \ref{subsec:Toy-example} and \ref{subsec:Ising-model}
both investigate empirically the impact of changing the quality of
estimates of $\mu_{\theta}$ and $\Sigma_{\theta}$ on estimates of
the SL. We then conclude with a discussion in section \ref{sec:Conclusions}.

\subsection{Bootstrapped synthetic likelihood\label{subsec:Bootstrapped-synthetic-likelihoo}}

In this section we introduce the bootstrap, and describe how it may
be used within SL. Our notation is the same as in section \ref{sec:Introduction},
except that we are now more precise about the distributions of each
quantity. The most common use of the the bootstrap is to estimate
the variance $\mathbb{V}_{Q \left(P\right)}\left[\theta_{N}\left(P_{N}\right)\right]$
(or some other property) of the sampling distribution $Q$ of
an estimator $\theta_{N}\left(P_{N}\right)$ of some population value
$\theta\left(P\right)$ based on data $y$ (with empirical distribution
$P_{N}$) from some unknown population distribution $P$. The useful
result exploited by the bootstrap is that the variance $\mathbb{V}_{Q\left(P\right)}\left[\theta_{N}\left(P_{N}\right)\right]$
may be accurately approximated by $\mathbb{V}_{Q\left(P_{N}\right)}\left[\theta_{N}\left(P_{N}\right)\right]$:
i.e. we may approximate the variance of $\theta_{N}$ by using the
empirical distribution of the data in place of the true population
distribution.

\subsubsection{Using the bootstrap to approximate $\Sigma_{\theta}$\label{subsec:Using-the-bootstrap}}

In the SL context we may exploit this idea since we wish to estimate
the variance $\Sigma_{\theta}=\mathbb{V}_{f_{\theta}\left(L_{\theta}\right)}\left[S_{N}\right]$
(to plug into the SL approximation) of an estimator $S_{N} = S_{N}\left(L_{N,\theta}\right)$
of $S\left(L_{\theta}\right)$. Here $L_{N,\theta}=\frac{1}{N}\sum_{i=1}^{N}\delta_{x_{i}}$
is the empirical distribution of a sample $x$ from $L_{\theta}$.
Using the bootstrap, we may approximate $\Sigma_{\theta}$ by $\varSigma_{\theta}=\mathbb{V}_{f_{\theta}\left(L_{N,\theta}\right)}\left[S_{N} \right]$. 

When using SL, it is possible to simulate multiple ($M$) samples
from $L_{\theta}$, thus we introduce an additional superscript $m$
into all quantities that depend on a sample $x^{(m)}$ from $L_{\theta}$.
To improve our approximation of $\Sigma_{\theta}$ we take the sample
average of multiple approximations $\varSigma_{\theta}^{(m)}=\mathbb{V}_{f_{\theta}\left(L_{N,\theta}^{(m)}\right)}\left[S_{N}^{(m)}\right]$,
yielding the approximation $\Sigma_{\theta}^{\text{boot}}=\frac{1}{M}\sum_{m=1}^{M}\varSigma_{\theta}^{(m)}$.
Standard results on the consistency of bootstrap estimates give us
that each $\varSigma_{\theta}^{(m)}\rightarrow\Sigma_{\theta}$ as
$N\rightarrow\infty$ (see \citet{Horowitz2001} for details). For
any $N$, the average $\Sigma_{\theta}^{\text{boot}}$ yields a lower
variance estimator than the individual $\varSigma_{\theta}^{(m)}$.

The quantities $\left\{ \varSigma_{\theta}^{(m)}\right\} _{m=1}^{M}$
are not available analytically, but may be estimated via resampling the single simulation $x^{(m)}$ from $L_{\theta}$. $R$ resamples
$\left\{ x^{(m,r)}\right\} _{r=1}^{R}$ from $L_{N,\theta}^{(m)}$
yield the Monte Carlo estimate
\[
\hat{\varSigma}_{\theta}^{(m)}=\hat{\mathbb{V}}_{f_{\theta}\left(L_{N,\theta}^{(m)}\right)}\left[S_{N}^{(m,1:R)}\right],
\]
where $S_{N}^{(m,r)}\sim f_{\theta}\left(L_{N,\theta}^{(m)}\right)$
is the summary statistic vector found from $x^{(m,r)}$, with the
sample variance $\hat{\mathbb{V}}_{f_{\theta}\left(L_{N,\theta}^{(m)}\right)}$ being taken over the $R$ summary statistic vectors.
Let $\hat{\Sigma}_{\theta}^{\text{boot}}=\frac{1}{M}\sum_{m=1}^{M}\hat{\varSigma}_{\theta}^{(m)}$,
where we for simplicity of notation we have omitted the dependence
of $\hat{\Sigma}_{\theta}^{\text{boot}}$ on $M$ and $R$.

Making computational savings through using bootstrapped SL (B-SL)
compared to standard SL requires that resampling a single simulation
$x$ is cheaper than simulating from the likelihood. This is the case
for most applications, but we may make further computational savings
when the B-SL approximation is used when $M$ is large, or when a
large number of $\theta$ are used as is the case when this approximation
is embedded within Monte Carlo methods. For i.i.d. data, a single
resample involves sampling $N$ times without replacement from $\left\{ 1,...,N\right\} $,
thus for $R$ resamples, $R\times N$ samples from $\left\{ 1,...,N\right\} $
are required; we denote such a sample by the matrix of indices $u=\left[u_{r,i}\right]_{r=1:R,i=1:N}$.
We may make a computational saving by reusing this same index matrix
for resampling every different simulation from the likelihood (i.e.
for every value of $m$ for every $\theta$).

Algorithm \ref{alg:boot} summarises our proposed procedure for estimating $\Sigma_{\theta}$. We will observe empirically in section \ref{sec:Empirical-results}
that, compared to standard SL, this approach significantly reduces
the variance of likelihood estimates without introducing noticeable
bias.

\begin{algorithm}
\begin{algorithmic}
\For {$m= 1:M$}
    \State $x^{(m)} \sim L_{\theta}$
    \For {$r= 1:R$}
        \State $x^{(m,r)} \sim L_{N,\theta}^{(m)}$
        \State Find $S_{N}^{(m,r)}$ from $x^{(m,r)}$.
    \EndFor
    \State $\hat{\varSigma}_{\theta}^{(m)}=\hat{\mathbb{V}}_{f_{N,\theta}\left(L_{N,\theta}^{(m)}\right)}\left[S_{N}^{(m,1:R)}\right]$.
\EndFor
\State Calculate $\hat{\Sigma}_{\theta}^{\text{boot}}=\frac{1}{M}\sum_{m=1}^{M}\hat{\varSigma}_{\theta}^{(m)}$.
\end{algorithmic}
\caption{The proposed bootstrap approximation to $\Sigma_{\theta}$.\label{alg:boot}}
\end{algorithm}

\subsubsection{Bootstrapped approximate Bayesian computation}

One may also consider using this approach in ABC, where the bootstrap
is used to obtain lower variance estimates of the ABC likelihood. For each sample $x^{\left(m\right)}$, the ABC kernel (often the uniform kernel with bandwidth $\epsilon$) is evaluated at the summary statistic $S_{N}^{(m,r)}$ of each resample, and the average of these results is taken to give the estimated likelihood for sample $m$. The bootstrapped ABC (B-ABC) likelihood is then the average of the estimated likelihoods for each $m$.
However, in this case we find that the bootstrapped estimates introduce
a significant bias into likelihood estimates. Specifically, we find
that a bootstrapped ABC (B-ABC) likelihood for some tolerance $\epsilon_{1}$
resembles the standard ABC likelihood for some $\epsilon_{2}>\epsilon_{1}$.
It is not clear in general whether lower variance estimates would
be achieved by using the standard ABC likelihood estimate for $\epsilon_{2}$,
or by using a B-ABC likelihood estimate for $\epsilon_{1}$. Empirical
results (section \ref{subsec:Lotka-Volterra-model}) suggest that
the bootstrapped approach can outperform the standard approach, but
need not always be the case. A significant drawback of the bootstrapped
approach is that the posterior will not converge to the true posterior
as the tolerance decreases to zero. Additionally we note that there
are alternative methods for achieving lower variance likelihood estimates
in ABC, such as \citet{Prangle2016}.

\subsubsection{Temporal and spatial models\label{subsec:Temporal-and-spatial}}

B-SL may also be used in situations outside of i.i.d. data, in any
situation where a bootstrapping procedure has been defined (the review
papers \citet{Horowitz2001} and \citet{Kreiss2012} give an overview
of the literature). Here we discuss the use of the block bootstrap,
which was introduced for bootstrapping stationary time series by \citet{Kunsch1989}.
Instead of resampling data independently, the block bootstrap instead
resamples blocks of data. These blocks are chosen to be sufficiently
large such that they retain the short range dependence structure of
the data, so that a resampled time series constructed by concatenating
resampled blocks has similar statistical properties to a real sample.
Below we outline the scheme that is used in this paper; others are
also possible (see \citet{Kreiss2012} for a review).

Suppose that $y_{1:N}$ is time indexed data, and that $x_{1:N}^{(m)}$
is a time series sampled from $l_{\theta}$. In the block bootstrap,
using a block of length $B$ (for simplicity we consider the case
where $B$ is a divisor of $N$), we first construct a set of overlapping
blocks of indices of the variables
\begin{equation}
\mathcal{B}=\left\{ (1:B),(2:B+1),...,(N-B+1:N)\right\} .\label{eq:index_set}
\end{equation}
Then a resample $x_{1:N}^{(m,r)}$ from $x_{1:N}^{(m)}$ consists
of $N/B$ concatenated blocks whose indices are sampled with replacement
from $\mathcal{B}$. The summary statistics of $R$ resamples $\left\{ x_{1:N}^{(m,r)}\right\} _{r=1}^{R}$
may then be computed, followed by calculating the sample variance
of these statistics over the $R$ resamples. This procedure is repeated
for each sample $m$, with the approximation to $\Sigma_{\theta}$
taken to be the sample mean of the variance estimate for each $m$.
As for the i.i.d. case, the resampling indices may be the same for
every $m$ and every $\theta$: again the indices may be stored in
a matrix $u=\left[u_{r,i}\right]_{r=1:R,i=1:N}$, where in this case
each row is given by concatenating the blocks of indices sampled from
$\mathcal{B}$.

In this paper we also study the case of stationary spatial models,
where the variables are arranged on a regular two-dimensional grid.
In this case we use the analogous scheme, where each resample is constructed
of randomly selected sub-tiles of the sample $x^{(m)}$. Again the
indices of the sub-tiles may be the same for every $m$ and $\theta$.

In both temporal and spatial models, for many summary statistics there
is an additional computational saving that is possible when $R$ is
large. We focus on the temporal case for simplicity. Suppose that
the summary statistic for a resampled time series may be computed
directly from corresponding statistics computed from its constituent
blocks. For example, the sample mean $\frac{1}{N}\sum_{i=1}^{N}x_{i}^{(m)}$
of a time series $x_{1:N}$ is given by $\frac{1}{N}\sum_{b=1}^{N-B+1}n_{k}B\left(\frac{1}{B}\sum_{k=1}^{B}x_{s_{b}+k-1}^{(m)}\right),$
where $s_{b}$ is the index at the beginning of the $b$th block,
the expression in the brackets is the sample mean of the $b$th block,
and $n_{k}$ is the number of times block $k$ appears in the resample.
In such a case, for each sampled time series $x^{(m)}$, rather than
compute the statistic for each resampled time series, it may be cheaper
to compute the statistic for each block then combine them. To see
this, consider the case where the cost of computing the statistic
is linear in the length $N$ of the time series. Here the cost of
computing the statistic for each resampled time series is $O\left(RN\right)$,
whereas when the statistics may be precomputed for each block the
cost is $O\left(\left(B+R\right)\left(N-B+1\right)\right)$. Therefore
as $R$ grows, the latter scheme offers a computational saving.

\subsection{Synthetic likelihood with the bag of little bootstraps\label{subsec:Synthetic-likelihood-with}}

For i.i.d. or stationary models, in the case where $N$ is large,
we may reduce the cost of B-SL through using the ``bag of little
bootstraps'' introduced by \citet{Kleiner2014}. This approach avoids
the simulation of data sets of size $N$, and instead constructs approximations
based on subsamples of size $n$, where $n<N$. We begin by outlining
the method mathematically, extending the notation of section \ref{subsec:Using-the-bootstrap}. We use the notation $l_\theta$ to denote the likelihood of a dataset of size $n$ (as opposed to previously, where $L_\theta$ is used for data of size $N$).

Recall that we wish to estimate the variance $\Sigma_{\theta}=\mathbb{V}_{f_{N,\theta}\left(L_{\theta}\right)}\left[S_{N}\right]$
of $S_{N}$. Using the bootstrap,
we approximated $\Sigma_{\theta}$ by $\varSigma_{\theta}=\mathbb{V}_{f_{N,\theta}\left(L_{N,\theta}\right)}\left[S_{N}\right]$.
The BLB instead uses the approximation $\varSigma_{n,\theta}=\mathbb{V}_{f_{N,\theta}\left(l_{n,\theta}\right)}\left[S_{N}\right]$
where $l_{n,\theta}=\frac{1}{n}\sum_{i=1}^{n}\delta_{x_{i}}$ is the
empirical distribution of a sample $x$ of size $n$ from $l_{\theta}$. In this
case, the empirical distribution used in place of $l_{\theta}$ only
requires data of size $n$ to be simulated from the likelihood, but
the resamples from $l_{n,\theta}$ are of size $N$. If $n=N$, this
method is the same as the standard bootstrap, with $\varSigma_{\theta}=\varSigma_{N,\theta}$.
The computational saving of the BLB arises since in practice, the
resamples of size $N$ need not actually be constructed: all that
needs to be stored are counts of the number of times each point in
the subsample is used in the resample, then all of the calculations
may be based on the subsample and these counts.

As previously, we simulate multiple ($M$) samples (this time of size
$n$) from $l_{\theta}$, and average over the approximations $\varSigma_{n,\theta}^{(m)}$
given by each sample yielding the approximation $\Sigma_{n,\theta}^{\text{blb}}=\frac{1}{M}\sum_{m=1}^{M}\varSigma_{n,\theta}^{(m)}$.
This procedure differs slightly to the BLB as introduced in \citet{Kleiner2014}, where there is
only a single dataset available. In such a case, the multiple simulations
used are $M$ subsets (simulated without replacement) of the data.
When used in SL (we refer to this approach as BLB-SL), we instead
simulate each ``subset'' independently from the likelihood, enabling
us to completely avoid the simulation of data of size $N$. For the
standard BLB, $\Sigma_{n,\theta}^{\text{blb}}\rightarrow\Sigma_{\theta}$
as $N\rightarrow\infty$ for any sequence $n\rightarrow\infty$ and
any fixed $M$, with convergence at the same rate as the bootstrap
(see \citet{Kleiner2014} for a precise statement of these results).
Importantly these results hold for $n\ll N$, giving the promise of
significant computational savings.

The BLB may be combined with the block bootstrap to be used for stationary
temporal and spatial data (previously considered for the temporal
case in \citet{Laptev2012}). Focussing on the temporal case, for
each $m$ we sample a time series of length $n$ from the likelihood.
From this we construct blocks of time series, using the index set
in equation \ref{eq:index_set} to index the blocks (using $B<n$).
We may think of each resampled time series of length $N$ as a concatenation
of $N/B$ blocks whose indices are sampled with replacement from $\mathcal{B}$,
although in practice this concatenation need not actually be constructed:
all that needs to be stored are the counts of the number of times
each block is selected for each resample. As previously, the indices
sampled from $\mathcal{B}$ may be the same for every $m$ and $\theta$.
In addition, if the summary statistic for a resample may be computed
directly from statistics of its constituent blocks, then for each
$m$ we may compute the statistic for each block then combine them.
The cost of this is $O\left(\left(B+R\right)\left(n-B+1\right)\right)$
which, crucially, is independent of $N$.

For estimating the variance $\Sigma_{\theta}$ fir SL, the BLB potentially offers a significant advantage over the
bootstrap, in that we need only simulate datasets of size $n$ rather
than size $N$, whilst maintaining accuracy. The empirical investigations in sections \ref{subsec:Toy-example}
and \ref{subsec:Ising-model} suggest that this potential is fulfilled
in practice.

\subsection{Regression for estimates of the expectation\label{subsec:Regression-for-estimates}}

The previous two sections focus exclusively on estimates of the variance
$\Sigma_{\theta}$. However, to use SL, we also need an estimate of
the mean $\mu_{\theta}$; in order that any computational saving is
made through using the bootstrap or BLB to estimate $\Sigma_{\theta}$,
the estimates of $\mu_{\theta}$ must not involve any further simulation
from the likelihood. Bootstrapping does not lead to lower variance
estimates of the mean, so other approaches are required.

When using the bootstrap to estimate $\Sigma_{\theta}$, we simulate
$M$ datasets of size $N$ from the likelihood. In this case we may
simply use the standard estimator used in SL in equation \ref{eq:sl_mu}.
When using the BLB, we simulate $M$ datasets of size $n$. For each
dataset of size $n$, we may estimate $\mu_{\theta}$ by simulating
a dataset of size $N$ from the corresponding empirical distribution
$l_{n,\theta}^{(m)}$ then again use the estimator in equation \ref{eq:sl_mu}.
Or alternatively, for statistics that satisfy a law of large numbers
in the size of the data (including all of those used in sections \ref{subsec:Toy-example}
and \ref{subsec:Lotka-Volterra-model}) it is sufficient (and introduces
less variance) to calculate the statistic based on $n$ data points
rather than $N$, then to simply use these raw estimates within equation
\ref{eq:sl_mu}. Section \ref{subsec:Ising-model} uses the same idea,
except that in this case the statistic needs to be scaled by the size
of the data, since it is a sum rather than an average. Estimates of
$\mu_{\theta}$ constructed in this way, based on only $n$ data points,
have a larger variance than those from $N$ data points. We expect
this increased variance to increase the variance of SL estimates,
but a further concern (borne out in practice) is that it also increases
the bias of SL estimates. \citet{Price2017} use an identity from
\citet{Ghurye1969} to correct for bias in the Gaussian estimator
resulting from errors in the expectation estimate, but this identity
is not applicable when using subsampling.

In section \ref{sec:Empirical-results} we observe empirically the
bias introduced through using high variance estimates of $\mu_{\theta}$.
The variance of these estimates may be reduced using regression, an
idea used in several previous papers. \citet{Meeds2014a,Sherlock2017}
describe MCMC schemes for estimating the likelihood as the MCMC algorithm
runs. In both of these methods, at each new value of $\theta$ visited
by the MCMC, random variables $x$ are simulated in order to estimate
the likelihood (in \citet{Meeds2014a} these are the simulations from
$l_{\theta}$ used in equations \ref{eq:sl_mu} and \ref{eq:sl_sigma}).
The likelihood regression then makes use of the entire history of
$\left(\theta,x\right)$ pairs built up as the MCMC runs. One weakness
of the approach in \citet{Sherlock2017} is that since the tails of
the posterior are visited infrequently, the regression estimates of
the likelihood in these regions has a higher variance. \citet{Meeds2014a}
address this issue by, where needed, actively acquiring additional
points in order to lower the variance of the regression estimate.
Also related is \citet{Moores2015} in which a preprocessing stage
is used to train a regression, which in this case is used to smooth
out the effect of using a finite value of $M$ in SL. In the current
paper we instead use the approach in \citet{Everitt2017c}, in which
a history of $\left(\theta,x\right)$ pairs is built up as an SMC
algorithm runs, where low variance regression estimates in the tails
are naturally obtained through beginning the SMC with a heavy tailed
proposal. For simplicity we use a local linear regression within this
SMC algorithm. In contrast to \citet{Meeds2014a}, we only perform
this regression on $\mu_{\theta}$, since we use the bootstrap for
approximating $\Sigma_{\theta}$. As remarked at the beginning of section \ref{sec:Methodology},
this removes a restriction of the \citet{Meeds2014a} approach, where
the regression on $\Sigma_{\theta}$ necessitates diagonalising the
covariance matrix.

\citet{Everitt2017c} uses marginal SMC \citep{DelMoral2006c} to
infer the posterior distribution. This algorithm maintains a population
of weighted particles $\left\{ \left(\theta_{t}^{(p)},w_{t}^{(p)}\right)\right\} _{p=1}^{P}$
that for each $t$ approximate a target distribution $\pi_{t}\left(\theta\right)=p\left(\theta\right)\hat{f}_{\theta}^{\nu_{t}}\left(y\right)$,
where $\hat{f}_{\theta}^{\nu_{t}}$ is an estimated likelihood raised
to a power and $\nu_{t}$ moves from 0 to 1 as $t$ increases. At
each target the kernel $K_{t}$ is used to move each particle, then
update
\begin{equation}
\tilde{w}_{t}^{(p)}=\frac{p\left(\theta_{t}^{(p)}\right)\hat{f}_{\theta_{t}^{(p)}}^{\nu_{t}}\left(y\right)}{\sum_{r=1}^{P}w_{t-1}^{(r)}K_{t}\left(\theta_{t}^{(p)}\mid\theta_{t-1}^{(r)}\right)}\label{eq:pmc_weight}
\end{equation}
is used at target $t$ to calculate unnormalised weights $\tilde{w}_{t}^{(p)}$
for the particles, which are normalised to give $w_{t}^{(p)}$. The
particles are then resampled. This approach has the advantage over
other SMC methods when using estimated likelihoods that bias in the
likelihood estimates does not accumulate as the algorithm runs. It
is particularly appropriate for the case where $\hat{f}_{\theta}$
is an estimated synthetic likelihood, since the estimates $\hat{\mu}_{\theta}$
at each $\theta$ may be stored, and used to fit regression models
to improve estimates of $\mu_{\theta}$ at future iterations. The
annealed sequence of distributions, which are heavier tailed than
the posterior in earlier iterations, leads to a useful set of estimates
to use in the regression. \citet{Everitt2017c} uses a similar approach
for doubly intractable distributions, and describes how (similar to
\citet{Sherlock2017}), for any particle $\theta_{t}^{(p)}$, to use
a KD-tree \citep{Bentley1975} to efficiently locate nearby previously
values of $\theta$ that may be used in the regression. Algorithm
\ref{alg:SMC} outlines our approach in full, in which a local linear regression
is used to estimate $\mu_{\theta_{t}^{(p)}}$. In this algorithm, each simulated $x$ has an additional superscript: the first refers to the index of the particle; the second to the index of the data simulated from $l_{\theta}$; the third (if present) to the index of the resample. As discussed in \citet{Everitt2017c},
this method is limited to applications where $\theta$ is of low to
moderate dimension.

\begin{algorithm}
\begin{algorithmic}
\For {$p= 1:P$}
    \State $\theta^{(p)}_0 \sim p\left( \cdot \right)$
    \For {$m= 1:M$}
        \State $x^{(p,m)}_{0} \sim f\left( \cdot \mid \theta^{(p)}_0 \right)$
     \EndFor
\EndFor
\State $t=0$.
\For {$t=0:T-1$}
    \For {$p= 1:P$}
        \State $\theta^{(p)}_{t+1} \sim K_{t+1}\left( \cdot \mid \theta^{(p)}_{t} \right)$
        \State $x^{(p)}_{t+1} \sim l_{\theta^{(p)}_{t+1}}\left( \cdot \right)$
    \EndFor
    \For {$p= 1:P$}
        \State Find the approximation $\mu^{\text{pred}}_{\theta^{(p)}_{t}}$ to $\mu_{\theta^{(p)}_{t}}$: the predicted value at $\theta^{(p)}_{t}$ from the local linear regression of the raw estimates of $\mu_{\theta}$ on $\theta$, fitted to the nearest $L$ points to $\theta^{(p)}_{t}$.
        \State Find the variance approximation $\Sigma_{n,\theta^{(p)}_{t}}^{\text{blb}}$ using BLB.
        \State $\tilde{w}_{t+1}^{(i)}=\frac{p\left(\theta_{t+1}^{(p)}\right) \left( \mathcal{N}\left( S(y) \mid \mu^{\text{pred}}_{\theta^{(p)}_{t}},\Sigma_{n,\theta^{(p)}_{t}}^{\text{blb}}\right) \right)^{\nu_{t+1}} }{ \sum_{q=1}^{P}w_{t}^{(q)}K_{t+1}\left(\theta_{t+1}^{(p)}\mid\theta_{t}^{(q)}\right)}$
    \EndFor
    \State Normalise $\left\{ \tilde{w}_{t+1} \right\}_{i=1}^N$ to give normalised weights $\left\{ w_{t+1} \right\}_{i=1}^N$.
    \State Resample.
\EndFor
\end{algorithmic}
\caption{SMC with BLB-SL using $M=1$.\label{alg:SMC}}
\end{algorithm}

\section{Empirical results\label{sec:Empirical-results}}

This section analyses the new methods empirically, through their application to i.i.d., temporal and spatial data.

\subsection{Toy example\label{subsec:Toy-example}}

In this section we explore the behaviour of B-SL, BLB-SL and ABC on
a toy model. We examine the bias and variance of likelihood estimates,
and the efficiency of Monte Carlo estimates when these approximate
likelihoods are used in MCMC algorithms. In this section regression
is not used to improve estimates of $\mu_{\theta}$.

Let data $y$ be $10^{5}$ points simulated from a univariate Gaussian
distribution with mean 0 and precision 0.25. Choosing a $\Gamma\left(1,1\right)$
prior on the precision $\tau$, we study the performance of the proposed
methods in estimating statistics of the posterior on the precision
$\tau$. In this example, the posterior is known by conjugacy. In
addition, the sample standard deviation is a sufficient statistic,
and its distribution conditional on $\tau$ is known analytically.
We may therefore estimate the error of estimates of statistics of
the posterior, and also of the likelihood approximations.

We ran MCMC algorithms with likelihoods given by SL, B-SL, BLB-SL,
ABC and B-ABC. All bootstrap algorithms used 100 resamples. For the
SL approaches, in order to distinguish the error resulting from using
a bootstrapped estimate of the variance from the error resulting from
using an estimated mean (which may have a high variance, particularly
in the BLB approaches), we run every SL approach with both the true
mean of the summary statistic for each $\tau$, and the estimated
mean. The proposal was taken to be normal with standard deviation
0.002. Each MCMC chain was started from the true posterior mean. For
our simulated $y$, the posterior mean and standard deviation are
respectively, to 3 s.f., $0.252$ and $1.13\times10^{-3}$. We ran
40 MCMC algorithms for each approximate likelihood, and report estimates
of the the bias, standard deviation, root mean squared error (RMSE)
of posterior mean and standard deviation estimates from the MCMC output.
In this toy example the likelihood is relatively easy to estimate,
and we find that for several of the approaches $M=10$ results in
MCMC algorithms that have similar autocorrelation to an MCMC algorithm
targeting the true posterior.

Figure \ref{fig:Estimated-bias,-standard} compares the efficiency
of posterior mean and standard deviation estimates from MCMC using
SL, B-SL, ABC and B-ABC estimates. The ABC algorithms used a Gaussian
distribution as the kernel in the ABC likelihood estimator, with standard
deviation $\epsilon=0.001$. We observe that standard SL performs
poorly for $M=2$, with the bias in estimates of the true summary
statistic likelihood leading to bias in the posterior mean and s.d..
The error in SL relative to ABC decreases as $M$ increases. For a
comparison of standard SL and ABC on a more challenging problem (where
there is a clear advantage to using SL), we refer the reader to section
\ref{subsec:Lotka-Volterra-model} (and also to \citet{Price2017}).

We now compare the performance of standard ABC and SL with their bootstrapped
versions. B-ABC has the advantage over ABC that for small values of
$M$ (when the likelihood variance is highest) it results in chains
of lower autocorrelation: for $M=1$, the mean estimated integrated
autocorrelation time (IAT) for the ABC chain is $\sim26$, compared
to $\sim9$ for B-ABC. However, as $M$ grows and the ABC estimates
of the likelihood improve, any advantage of B-ABC is negated, particularly
since it results in an overestimation of the posterior uncertainty,
clearly seen in figure \ref{fig:Estimated-bias-of}. B-SL exhibits
improved performance over SL in almost every case (and can be implemented
for $M=1$, where SL cannot). These comparisons are revisited in section
\ref{subsec:Lotka-Volterra-model} on a more challenging example.
The comparison of SL and B-SL with the case where the true mean is
used in place of the estimated mean suggests the potential of an approach
that combines the bootstrap estimates of $\Sigma_{\theta}$ with improved
estimates of $\mu_{\theta}$. We see that B-SL with the true value
of $\mu_{\theta}$ outperforms all other approaches: the results are
comparable to using an MCMC with the true likelihood.

\begin{multifig}
\shortfig{bias_mean}{
  scale=.35,
  caption=Estimated bias of posterior mean estimates.,
}
\shortfig{sd_mean}{
  scale=.35,
  caption=Estimated standard deviation of posterior mean estimates.,
 }
\shortfig{rmse_mean}{
  scale=.35,
  caption=Estimated RMSE of posterior mean estimates.,
 }
\shortfig{bias_sd}{
  scale=.35,
  caption=Estimated bias of posterior standard deviation estimates.,
  label=fig:Estimated-bias-of,
 }
\shortfig{sd_sd}{
  scale=.35,
  caption=Estimated standard deviation of posterior standard deviation estimates.,
 }
\shortfig{rmse_sd}{
  scale=.35,
  caption=Estimated RMSE of posterior standard deviation estimates.,
 }
 
 \caption{Estimated bias, standard deviation and RMSE from standard and bootstrapped
versions of SL and ABC.\label{fig:Estimated-bias,-standard}}

\end{multifig}

Figure \ref{fig:Estimated-bias,-standard-1} compares the efficiency
of posterior mean and standard deviation estimates from MCMC using
different BLB-SL approximations. We observe that the subsampling has
two large effects (both of which are increased by decreasing the size
of the subsample): the posterior s.d. is overestimated (see figure
\ref{fig:Estimated-bias-of-1}); and the variance of the estimates
is increased (due to an increased autocorrelation in the chains).
However, we observe that both of these effects are reduced dramatically
by using the true value of $\mu_{\theta}$ in the SL estimates. In
this situation, the autocorrelation in the MCMC chains and the errors
in posterior estimates are similar between B-SL and BLB-SL, no matter
the size of the subsample (for $n=10,000$, $1000$ or $100$). This
suggests great potential for the BLB approach, as long as accurate
estimates of $\mu_{\theta}$ may be obtained through other means.
Section \ref{subsec:Ising-model} illustrates that the regression
approach suggested in section \ref{subsec:Regression-for-estimates}
provides a way of achieving this.

\begin{multifig}
\shortfig{blb_bias_mean}{
  scale=.35,
  caption=Estimated bias of posterior mean estimates.,
}
\shortfig{blb_sd_mean}{
  scale=.35,
  caption=Estimated standard deviation of posterior mean estimates.,
 }
\shortfig{blb_rmse_mean}{
  scale=.35,
  caption=Estimated RMSE of posterior mean estimates.,
 }
\shortfig{blb_bias_sd}{
  scale=.35,
  caption=Estimated bias of posterior standard deviation estimates.,
  label=fig:Estimated-bias-of-1,
 }
\shortfig{blb_sd_sd}{
  scale=.35,
  caption=Estimated standard deviation of posterior standard deviation estimates.,
 }
\shortfig{blb_rmse_sd}{
  scale=.35,
  caption=Estimated RMSE of posterior standard deviation estimates.,
 }
 
\caption{Estimated bias, standard deviation and RMSE from BLB-SL samplers.\label{fig:Estimated-bias,-standard-1}}
 
\end{multifig}

\subsection{Lotka-Volterra model\label{subsec:Lotka-Volterra-model}}

\subsubsection{Introduction}

The Lotka-Volterra model is well-studied in the ABC literature.
The model is a stochastic Markov jump process that describes how the
number of individuals in two populations (one of predators, the other
of prey) change over time. We use the form of the model in \citet{Wilkinsona},
in which $X$ represent the number of predators and $Y$ the number
of prey. The following reactions may take place:
\begin{itemize}
\item A prey may be born, with rate $\theta_{1}Y$, increasing $Y$ by one.
\item The predator-prey interaction in which $X$ increases by one and $Y$
decreases by one, with rate $\theta_{2}XY$.
\item A predator may die, with rate $\theta_{3}X$, decreasing $X$ by one.
\end{itemize}
Figure \ref{fig:Data-simulated-from} shows the simulated data $y$
studied in this section: it consists of two oscillating time series:
one giving the size of the predator population, the other of the prey.
The simulation starts with initial populations $X=50$ and $Y=100$,
and including the initial values has 32 measurements for each series,
with the values of $X$ and $Y$ being recorded every 2 time units.
The model may be simulated exactly using the Gillespie algorithm \citep{Gillespie1977},
but it is not possible to evaluate its likelihood. We followed the
ABC approaches in \citet{Wilkinsona,Papamakarios2016}, using as summary
statistics a 9-dimensional vector composed of the mean, log variance
and first two autocorrelations of each time series, together with
the cross-correlation between them (scaled by dividing by the summary
statistic vector $T$ of the observed data).

\begin{figure}
\includegraphics[scale=0.2]{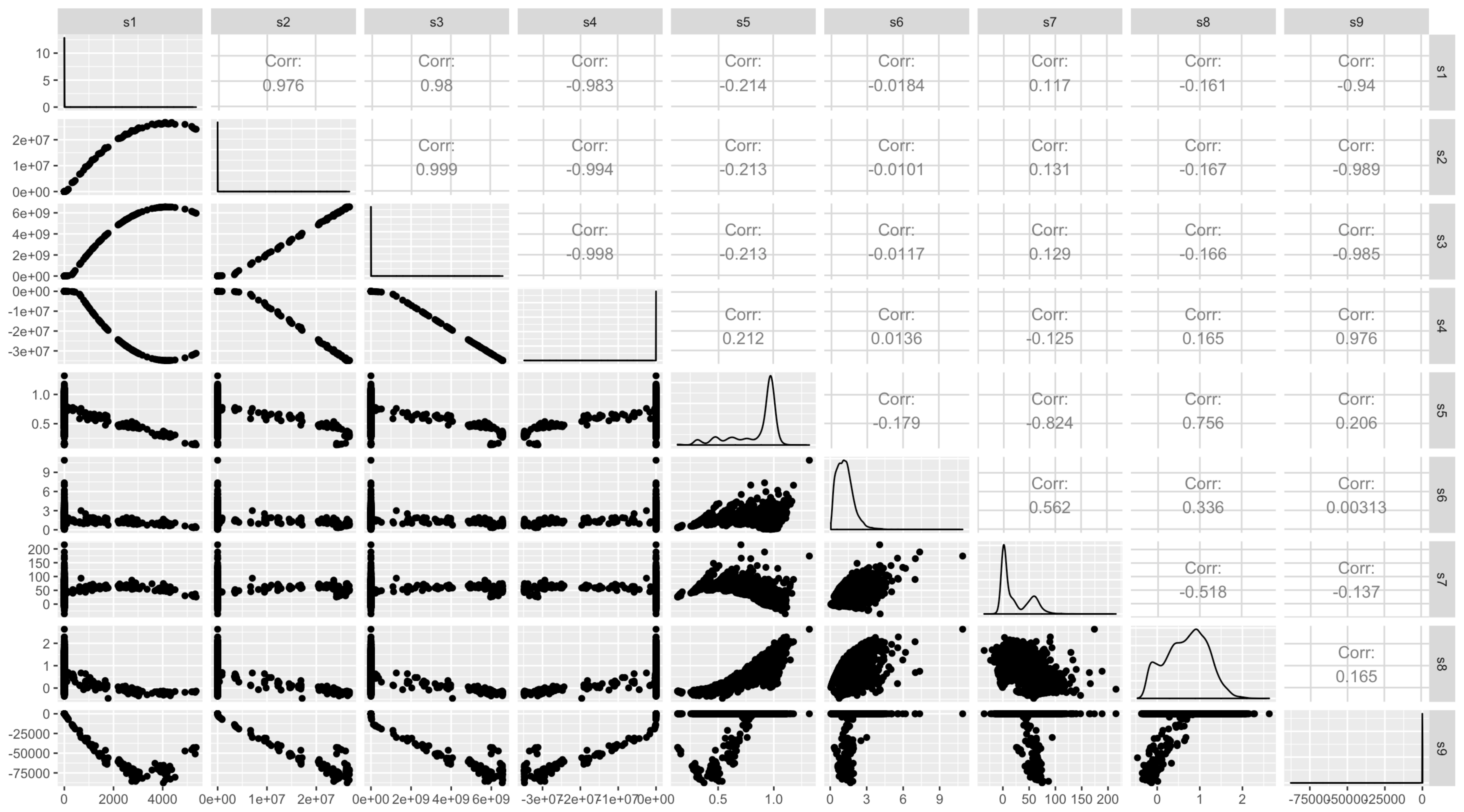}

\caption{Scatter plots and kernel density estimates from draws from the distribution
of the summary statistics conditional on the true parameters, and
estimated correlations between the statistics.\label{fig:Scatter-plots-of}}

\end{figure}

This model has a number of properties that provide a challenge to
our proposed approach:
\begin{enumerate}
\item The simulations from the model give temporal data, which requires
the use of the block bootstrap (as described in section \ref{subsec:Temporal-and-spatial}).
\item The simulations from the model cannot always be considered to be stationary
time series, since sometimes (more commonly for inappropriate parameters)
the sizes of the populations decreases to zero, or diverges towards
infinity. We do not treat such simulations differently to any other
simulation, thus our results help to illustrate how robust our approach
is to this situation.
\item The 9-dimensional summary statistics allow us to illustrate the performance
of our bootstrapping approach for estimating a covariance matrix.
\item The distribution of the summary statistics is not close to being Gaussian,
and there are complex dependencies between the statistics (see figure
\ref{fig:Scatter-plots-of}). The former point allows us to examine
the performance of SL when the Gaussian assumption is not satisfied
(as previously studied in \citet{Price2017}; the latter suggests
that the approach of \citet{Meeds2014a} may not be appropriate, since
they assume a diagonal covariance matrix.
\end{enumerate}

We compared the output of MCMC algorithms employing the approximate
likelihoods given by SL, B-SL, ABC and B-ABC. We ran each method with
several different choices of $M$, and the bootstrap algorithms used
$R=100$ resamples. The algorithms were run for $5\times10^{4}$ iterations,
and were initialised at the parameters $\theta_{1}=1$, $\theta_{2}=0.005$
and $\theta_{3}=0.6$ for which the data $y$ was simulated. The MCMC
proposal was a multivariate Gaussian with diagonal covariance matrix
whose diagonal is $\left(0.2^{2},0.001^{2},0.2^{2}\right)$, these
values being determined using pilot runs. Our prior followed \citet{Wilkinsona},
being uniform in the $\log$ domain
\[
p\left(\log\left(\theta\right)\right)\propto\prod_{i=1}^{3}\mathcal{U}\left(\log\left(\theta_{i}\right)\mid\text{lower}=-6,\text{upper}=2\right).
\]
The block bootstrap used blocks of length 8, so that each bootstrap
resample consists of 4 blocks. Each statistic of each resample is
calculated by combining the corresponding statistic of each of the
constituent blocks in the obvious way.

\subsubsection{Results}

Table \ref{tab:The-estimated-integrated} shows the mean (over the
three parameters) estimated IAT of each sampler, and figure \ref{fig:SL-and-ABC}
(b-f) shows kernel density estimates of the marginal posterior distributions
based on the MCMC samples. In comparing the results from standard
SL to standard ABC, we see that SL usually results in more efficient
MCMC samplers, and we do not see any clear indications that the Gaussian
assumption made in SL is problematic. The bootstrapped algorithms
do not appear to be adversely affected by any of the challenges described
in the previous section, giving similar posterior distributions to
the standard approaches. Further, the autocorrelation properties of
the MCMC chains from the bootstrapped algorithms are improved over
their standard counterparts. This provides further evidence for the
observation made in section \ref{subsec:Toy-example} that the bootstrapped
methods are useful when the corresponding standard estimates of the
likelihood have a high variance.

Bootstrapped SL consistently exhibits the best performance: figure
\ref{fig:Density-estimates-for} shows that the low IAT found for
SL when $M=5$ is not representative of the performance of the algorithm,
since the sampler is only exploring a region in the tails of the posterior.

\begin{table}
\begin{tabular}{|c|c|c|c|c|c|}
\hline 
Algorithm & $M=1$ & $M=2$ & $M=5$ & $M=10$ & $M=50$\tabularnewline
\hline 
\hline 
SL & N/A & \textcolor{red}{3749} & \textcolor{green}{273} & 4210 & 326\tabularnewline
\hline 
B-SL & \textcolor{green}{3543} & \textcolor{green}{1141} & 504 & \textcolor{green}{190} & \textcolor{green}{168}\tabularnewline
\hline 
ABC $\epsilon=0.1$ & 6006 & 2969 & \textcolor{red}{4813} & \textcolor{red}{4506} & \textcolor{red}{1974}\tabularnewline
\hline 
B-ABC $\epsilon=0.1$ & 5443 & 2874 & 1409 & 496 & 199\tabularnewline
\hline 
ABC $\epsilon=0.2$ & \textcolor{red}{13124} & 2923 & 2223 & 1416 & 427\tabularnewline
\hline 
\end{tabular}

\caption{The estimated integrated autocorrelation time (to 0 d.p.) of each
chain.\label{tab:The-estimated-integrated}}

\end{table}

\begin{multifig}
\shortfig{LV32_data}{
  scale=.55,
  caption=Data simulated from the Lotka-Volterra model.,
  label=fig:Data-simulated-from,
}
\shortfig{lv1_kde1}{
  scale=.55,
  caption=Density estimates for \M1.,
 }
\shortfig{lv1_kde2}{
  scale=.55,
  caption=Density estimates for \M2.,
 }
\shortfig{lv1_kde5}{
  scale=.55,
  caption=Density estimates for \M5.,
  label=fig:Density-estimates-for,
 }
\shortfig{lv1_kde10}{
  scale=.55,
  caption=Density estimates for \M10.,
 }
\shortfig{lv1_kde50}{
  scale=.55,
  caption=Density estimates for \M50.,
 }
 
\caption{SL and ABC algorithms applied to the Lotka-Volterra model.\label{fig:SL-and-ABC}}
 
\end{multifig}

\subsection{Ising model\label{subsec:Ising-model}}

Undirected graphical models, or Markov random fields (MRFs), have
previously been studied using ABC in a number papers, beginning with
\citet{Grelaud2009}. Such models have the form
\[
L_{\theta}\left(y\right)=\frac{\gamma_{\theta}\left(y\right)}{Z\left(\theta\right)},
\]
where $\gamma_{\theta}\left(y\right)$ is tractable, but the partition
function $Z\left(\theta\right)$ cannot, in practice, be evaluated
pointwise. \citet{Moller2006,Murray2006} pioneered the approach of
estimating $L_{\theta}$ at each $\theta$ using importance sampling
(known as auxiliary variable methods), embedded in an MCMC algorithm
to perform Bayesian inference on $\theta$. ABC may be used as an
alternative, but the likelihood estimates are typically high variance
in comparison with those from the \citet{Moller2006} approach (\citet{Everitt2017c}
establishes a connection between the two approaches). However, when
using a latent MRF model, ABC can be competitive with auxiliary variable
methods \citep{Everitt2017b}. Also, SL provides a lower variance
alternative to ABC that can be competitive with auxiliary variable
methods \citep{Moores2015,Everitt2017}. All of these previous approaches
require simulating from $L_{\theta}$, which for most MRFs needs to
be done approximately by using a run of MCMC with $L_{\theta}\left(\cdot\right)$
as the target distribution \citep{Caimo2011}. The use of MCMC introduces
an approximation, which is small as long the chain is run long enough
to have essentially forgotten its initial condition \citep{Everitt2012}.

In this section, we focus on the Ising model. This is a pairwise Markov
random field model on binary variables, each taking values in $\left\{ -1,1\right\} $.
Its distribution is given by
\[
L_{\theta}\left(y\right)\propto\exp\left(\theta\sum_{\left(i,j\right)\in\mathbf{N}}y_{i}y_{j}\right),
\]
where $\theta_{x}\in\mathbb{R}$, $x_{i}^{h}$ denotes the $i$th
random variable in $x^{h}$ and where $\mathbf{N}$ is a set that
defines pairs of nodes that are \textquotedblleft neighbours\textquotedblright .
We consider the case where the neighbourhood structure is given by
a regular 2-dimensional grid, using a first order model (so that variables
horizontally and vertically adjacent are neighbours) and toroidal
boundary conditions, and use a Gibbs sampler to simulate from $L_{\theta}\left(\cdot\right)$.
The mixing properties of the Gibbs sampler on Ising models are well
understood, and indicate a limitation of all of the approaches to
inference outlined above: namely that the approaches do not scale
to large MRFs. The mixing time of Gibbs samplers on Ising models on
2-d grids is at best polynomial in the number of rows in the grid
\citep{Lubetzky2012}. Therefore, as the size of the grid grows, in
addition to the number of single variable updates growing linearly
in the size of the grid, we expect to need to run the Gibbs sampler
for more iterations.

In this section we study data from a $1,000\times1,000$ Ising model
(so that $N=10^{6}$), generated with $\theta=0.3$, and compare results
from the exchange algorithm (an auxiliary variable MCMC approach introduced
in \citet{Murray2006}) and BLB-SL. In all cases, the Gibbs sampler
for simulating from the likelihood is burned in for 10 iterations.
The exchange algorithm was initialised at $\theta=0.298$ and run
for $1000$ iterations, using a normal proposal with standard deviation
$0.001$. For BLB-SL, the spatial block bootstrap was used (as described
in section \ref{subsec:Temporal-and-spatial}), with the size of the
subsample being either $100\times100$ or $50\times50$ (so that $n=10^{4}$
or $2,500$) and the block size being $50\times50$ or $25\times25$
(so that $B=2,500$ or $625$). The (sufficient) statistic $S\left(y\right)=\sum_{\left(i,j\right)\in\mathbf{N}}y_{i}y_{j}$
was used, and we took $M=1$. The SMC algorithm from section \ref{subsec:Regression-for-estimates}
was used, with $P=1,000$ particles and $T=10$ target distributions,
with $\nu_{t}=\left(t/T\right)^{2}$.  For each $\theta_{t}^{(p)}$
(the $p$th particle at the $t$th target) a sample $x_{t}^{(p)}$
of size $\sqrt{n} \times \sqrt{n}$ is simulated from $l_{\theta}\left(\cdot\right)$.
$\mu_{\theta}$ and $\Sigma_{\theta}$ was then approximated as follows.
\begin{itemize}
\item Calculate $\left(N/n\right)S_n$ (the statistic
rescaled from a grid of size $n$ to a grid of size $N$) as a ``raw''
estimate of $\mu_{\theta}$. Find the $C$ closest $\theta$ values
to $\theta_{t}^{(p)}$ (including $\theta_{t}^{(p)}$ itself)  and
perform a linear regression of $\left(N/n\right)S_n$
on $\theta_{t}^{(p)}$. Then, from the regression use the predicted
value of the response at $\theta_{t}^{(p)}$ as the estimate of $\mu_{\theta}$.
\item Compose $R=100$ resamples from $x_{t}^{(p)}$ by using the following
procedure
\begin{itemize}
\item Take (overlapping) blocks of size $\sqrt{B}\times \sqrt{B}$ from $x_{t}^{(p)}$
as described in the spatial block bootstrap: there are $(1+\sqrt{B}) \times (1+\sqrt{B})$
of these blocks in total. Compute the statistic for each block, denoting
it by $\mathcal{S}_b$ for the $b$th block.
\item Randomly compose a resample of size $\sqrt{N} \times \sqrt{N}$ by piecing together
$N/B$ blocks. The indices $\mathcal{B}$ of the blocks used in the
resample may be the same for all $p$ and $t$, thus can be generated
in a pre-processing step.
\end{itemize}
\item Compute the statistic for each resample, using for the $r$th resample
\[
S_{N}^{(r)}=\left(\frac{N}{N-\frac{N}{\sqrt{B}}}\right)\sum_{b\in\mathcal{B}} \mathcal{S}_b,
\]
where the rescaling accounts for the absence of edges between the
blocks. The sample variance of the $\left\{ S_{N}^{(r)}\right\} _{r=1}^{R}$
is then our approximation of $\Sigma_{\theta}$ from the block-BLB.
\end{itemize}
Figure \ref{fig:Density-estimates-from} shows the estimated posterior
distributions from the exchange algorithm, and runs of the BLB-SL
SMC method for different values of $C$ and $n$. We observe that
the posterior distribution from the exchange algorithm is very well
approximated by the posterior distribution from BLB-SL SMC when $C=100$
and $n=10,000$, with the posterior standard deviation being overestimated for smaller $C$. This is due to the increasing
variance estimates of the SL mean as $C$ decreases, as was
previously observed in section \ref{subsec:Toy-example}. In this example, the combination (through regression) of $C=100$ raw estimates
of $\mu_{\theta}$ is sufficient to reduce the variance sufficiently
that an accurate posterior results. This variance reduction, without the introduction of significant bias, is possible since the assumptions made in the regression are
appropriate. Figure \ref{fig:Raw-estimates-of} illustrates shows the raw estimates
of $\mu_{\theta}$ against $\theta$ in the region of the posterior, together
with the predicted regression with $C=100$ for each point. We also
find that using $C=200$ and $n=2,500$ yields a fairly accurate posterior,
indicating that our approach can be accurate even when the subsampling
ratio $N/n$ is quite large ($400$ in this case). Figure \ref{fig:The-ESS-over}
shows the effective sample size (ESS) over the iterations of the SMC
sampler for different values of $C$. As in section \ref{subsec:Toy-example}
we observe that the efficiency of the Monte Carlo method in which
the BLB-SL estimate is embedded decreases when the estimates of the
mean have higher variance. For comparison, the ESS for the exchange
algorithm was estimated at $98$ using the \texttt{LaplacesDemon}
package in R (whilst recalling that the ESS is defined differently
for importance sampling and MCMC algorithms).

\begin{multifig}
\shortfig{ising_density}{
  scale=.5,
  caption=Density estimates from each of the Monte Carlo algorithms.,
  label=fig:Density-estimates-from,
}
\shortfig{ising_ess}{
  scale=.5,
  caption=The ESS over SMC iterations for the BLB-SL algorithms.,
  label=fig:The-ESS-over,
 }
\shortfig{regression}{
  scale=.7,
  caption=Raw estimates of \mut (blue) and the locally linear regression
prediction (red) with \El100.,
  label=fig:Raw-estimates-of,
  align=c,
 }
 
 \caption{Results of applying BLB-SL to the Ising model.}

\end{multifig}

\section{Conclusions\label{sec:Conclusions}}

This paper introduces methodology for improving the performance of
SL in cases where a bootstrap may be used to estimate the variance
of the chosen statistics. Further, it provides a method for using
SL in ``tall data'' settings, where subsampling may be used such
that the cost of the sampling algorithm depends on the size $n$ of
the subsets rather than $N$, the size of the full data. In summary
\begin{itemize}
\item In situations in which the likelihood is difficult to estimate, bootstrap
approximations of the variance of a statistic result in lower variance
likelihood estimates, thus improve the efficiency of Monte Carlo methods
that use SL. Further, using the bootstrap only results in a small
bias. The same is true to an extent when using bootstrapped estimates
of the ABC likelihood.
\item Using the BLB to estimate the variance of a statistic has a similar
performance to using the bootstrap, paving the way for using subsampling
to estimate SLs. However, estimates of the mean of the statistic using
subsamples are too high variance to result in either accurate approximations
of the true posterior distribution, or low variance Monte Carlo algorithms.
\item When using the BLB, regression estimates of the statistic mean may
be used in order to reduce the variance sufficiently that an accurate
approximation to the true posterior is obtained, even for large values
of $N/n$. In this paper a local linear regression was used, but in
other models different regression techniques such as Gaussian processes
may be more appropriate.
\end{itemize}
The methods in this paper should be of use whenever a bootstrap method
is available to estimate the variance of the chosen statistics, with
the BLB being applicable to stationary models. It is possible that,
since it only requires the simulation of subsamples of size $n$,
BLB-SL may be useful in big data settings where ABC/SL would not usually
be applied. One further remark about the big data setting is that
often in such cases sophisticated Monte Carlo methods are not required
since the posterior is approximately Gaussian. This is not necessarily
the case in many models where ABC/SL might be used, since parameters
are often non-identifiable, leading to complex posterior distributions
no matter how much data is observed (see the differential equation
models in \citet{Maybank2017}, for example).

\bibliographystyle{/Users/Everitt/Dropbox/projects/bib/mychicago}
\bibliography{bootstrap_paper}

\end{document}